\documentclass{emulateapj}

\begin{document}

\shorttitle{Evolution of Globular Clusters}
\shortauthors{Kalirai et~al.}

\title{A Glimpse into the Past: The Recent Evolution of Globular Clusters\altaffilmark{1,2}}

\author{
Jasonjot S. Kalirai\altaffilmark{3,5}, 
Jay Strader\altaffilmark{4,5}
Jay Anderson\altaffilmark{6}, and
Harvey B. Richer\altaffilmark{7}
}
\altaffiltext{1} {Based on observations with the NASA/ESA Hubble Space Telescope, obtained at 
the Space Telescope Science Institute, which is operated by the Association of Universities 
for Research in Astronomy, Inc., under NASA contract NAS5-26555.  These observations are 
associated with proposal GO-10424.}
\altaffiltext{2} {Based on observations obtained at the Gemini Observatory, which is operated by the
Association of Universities for Research in Astronomy, Inc., under a cooperative agreement
with the NSF on behalf of the Gemini partnership: the National Science Foundation (United
States), the Particle Physics and Astronomy Research Council (United Kingdom), the
National Research Council (Canada), CONICYT (Chile), the Australian Research Council
(Australia), CNPq (Brazil) and CONICET (Argentina).}
\altaffiltext{3}{University of California Observatories, University of 
California at Santa Cruz, Santa Cruz CA, 95060; jkalirai@ucolick.org}
\altaffiltext{4}{Harvard-Smithsonian Center for Astrophysics, Cambridge MA, 02138; 
jstrader@cfa.harvard.edu}
\altaffiltext{5}{Hubble Fellow}
\altaffiltext{6}{Space Telescope Science Institute, Baltimore MD, 21218; jayander@stsci.edu}
\altaffiltext{7}{Department of Physics and Astronomy, University of British Columbia, 
Vancouver, BC, Canada, V6T~1Z1; richer@astro.ubc.ca}


\begin{abstract}

We present the serendipitous discovery of 195 extragalactic globular 
clusters (GCs) in one of the deepest optical images ever obtained, a 126 orbit 
{\it Hubble Space Telescope} ({\it HST}) Advanced Camera for Surveys (ACS) 
imaging study of the nearby Galactic GC NGC~6397.  The 
distant GCs are all found surrounding a bright elliptical 
galaxy in the field, and are among the faintest objects detected in the image, 
with magnitudes 26 $\lesssim F814W \lesssim$ 30.  We measure the redshift of the 
parent elliptical galaxy, using the Gemini Multi-Object Spectrograph (GMOS) on 
Gemini South, to be $z$ = 0.089 (375~Mpc).  This galaxy, and its associated 
clusters, therefore ranks as one of the most distant such systems discovered 
to date.  The measured light from these clusters was emitted 1.2~Gyr ago 
(the lookback time) and therefore the optical properties hold clues for 
understanding the evolution of GCs over the past Gyr.  
We measure the color function of the bright GCs and find that both a 
blue and red population exist, and that the colors of each sub-population 
are redder than GCs in local elliptical galaxies of comparable 
luminosity.  For the blue clusters, the observed color difference from $z = 0.089$ 
to today is only slightly larger than predictions from stellar evolution (e.g., 
changes in the luminosity and color of the main-sequence turnoff and the 
morphology of the horizontal branch).  A larger color difference is found in the 
red clusters, possibly suggesting that they are very metal-rich and/or significantly 
younger than 12~Gyr.

\end{abstract}


\section{Introduction} \label{introduction}

The formation of globular star clusters (GCs) is closely linked to the 
evolutionary histories of their parent galaxies.  Major epochs of star 
formation in isolated galaxies as well as interactions with other 
galaxies lead to subsequent spikes in the number of GCs 
\citep{schweizer87,ashman98,harris01,brodie06}.  Indeed, 
it appears that the processes involved in galaxy formation imprint a 
signature on the clusters as reflected by the similar properties of 
the host galaxy and its associated clusters (e.g., age and metallicity).  
The prevalence of a large number of GCs around all 
massive galaxies, as well as their optically bright luminosities ($-$10 $ 
\lesssim M_{V} \lesssim -$5), makes them excellent tracers of galaxy 
formation processes.


\begin{figure*}
\epsscale{0.95}
\plotone{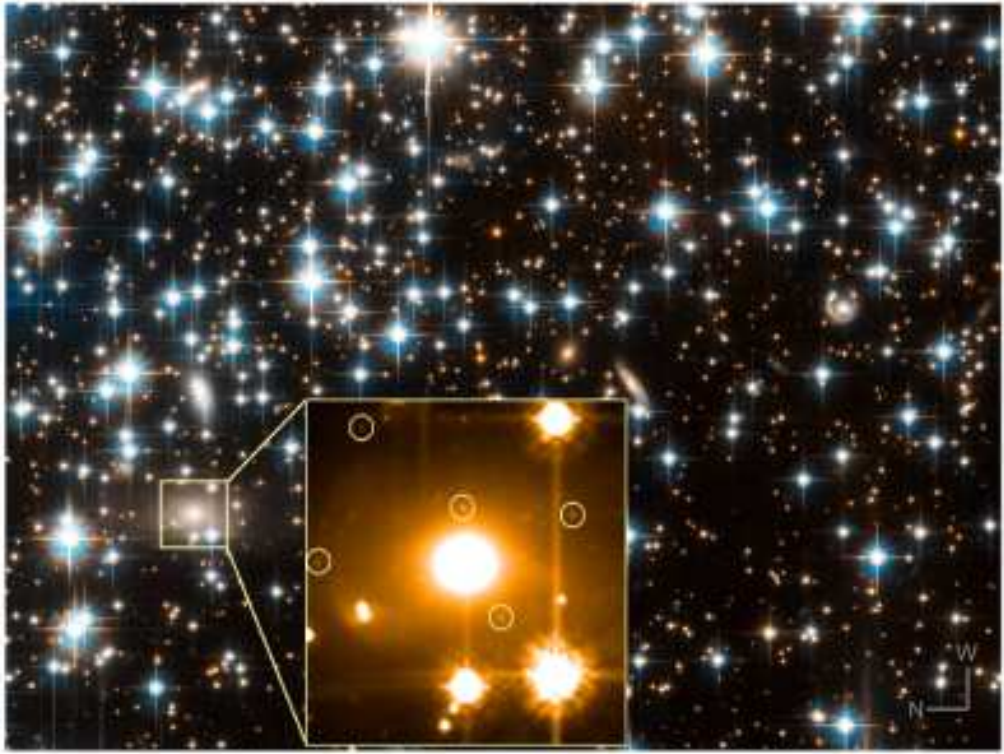}
\figcaption{A 3.4$'$ $\times$ 1.7$'$ section of the 126 orbit 
{\it HST} exposure of NGC~6397 (this is approximately one-half of 
the total ACS field of view).  The strong gradient across the image 
from top-left to bottom-right reflects the decreasing radial density 
of NGC~6397 stars (the cluster center is located $\sim$6.5 arcminutes 
to the N-W of our field).  The inset shows a closer look (8$''$ $\times$ 
8$''$ -- 15.9 $\times$ 15.9~kpc at the distance of the elliptical galaxy) 
of the large elliptical galaxy at 
image coordinates ($x$, $y$) = (789.9, 750.9) pixels.  The point source 
enhancement surrounding the elliptical galaxy is very high and suggests 
that these objects are GCs themselves, orbiting this 
galaxy.  Over the area of the inset, only one cluster white dwarf is 
expected to contaminate our sample.  Five of the GCs 
are highlighted with small circles in the inset.  Higher resolution 
version of this figure is available at http://www.ucolick.org/$\sim$jkalirai/DistantGlobs/.
Image credit: NASA, ESA, H.\ Richer (UBC), J.\ Kalirai (UCSC).
\label{fig:image}}
\end{figure*}



To date, GCs have been detected in hundreds of nearby 
galaxies (see Table~1 in Brodie \& Strader 2006 for those systems with 
accurate photometry).  However, direct age measurements from resolved 
main-sequence turnoff fitting are only possible for the nearest clusters 
in the Milky Way, LMC, SMC, and M31 (e.g., see the recent study of SKHB~312 
by Brown et~al.\ 2004).  Properties of more distant systems are typically 
constrained with integrated light photometry and, where possible, 
spectroscopy.  Such work has demonstrated that the luminosity and color 
distributions of GCs in most galaxies are remarkably similar (e.g., 
Harris \& Racine 1979), with most populations exhibiting a bimodal color 
distribution reflecting old metal-poor and metal-rich subpopulations.  
Whether these populations reflect GCs that formed at two different 
times (e.g., Forbes et~al.\ 1997) is still debated.  Unfortunately, 
directly probing the properties of these clusters at earlier times is a 
difficult task.  For example, at the distance of the nearby Coma cluster, 
a typical GC with $M_{V}$ = $-$7.5 has an apparent 
magnitude of $V$ = 27.5.

In {\it HST} Cycle 13, our team (PI: Richer - GO-10424) was granted 126 
orbits of ACS time in a single pointing to explore the white dwarf cooling 
sequence, faint main-sequence, and space motion of the Milky Way 
core-collapsed GC NGC~6397 
\citep{richer06,hansen07,kalirai07a}.  The resulting data set represents 
one of the deepest optical images ever obtained in astronomy, with a 50\% 
(25\%) completeness limit of $F814W$ = 28 ($F814W$ = 28.75) and a 
photometric error of $\sigma$ = 0.3 at $F814W$ = 28.0 and $F606W$ = 29.25.  
In this {\it Letter}, we present the discovery of a distant, background 
galaxy in this data set that harbors its own GC population.  
We measure the redshift of the galaxy to be $z$ = 0.089, and therefore this 
system resides $\sim$375~Mpc from the Milky Way at a lookback time of 
1.2~Gyr.  The color function of the GCs in this galaxy is 
found to show differences when compared to local samples suggesting the 
possible detection of evolutionary changes in GC properties 
over the past Gyr.

\section{Imaging and Spectroscopic Observations} \label{observations}

The {\it HST}/ACS imaging observations of NGC~6397 from GO-10424 have been 
described in several papers (e.g., Richer et~al.\ 2006; Hansen et~al.\ 2007; 
Kalirai et~al.\ 2007a; Anderson et~al.\ 2008).  To summarize, we 
obtained 252 images in $F814W$ and 126 in $F606W$, for a total 
integration time of 126 orbits in a single pointing centered at 
$\alpha_{\rm J2000}$ = 17:41:03, $\delta_{\rm J2000}$ = $-$53:44:21.  
Sources detected at the same position on 90 of the 252 $F814W$ 
images were shown to be a 3$\sigma$ detection (from simulations) 
and cleaned to remove false detections in the wings of bright stars, those 
caused by intersecting diffraction spikes, and objects morphologically 
inconsistent with stars (e.g., galaxies and cosmic rays).  PSF 
photometry on these sources was performed as described in 
\cite{anderson08}.  

Our expectation was that the final cleaned catalog of $\sim$8000 objects 
from this analysis represented stars on the image.  However, a 
spatial plot of the faint-blue sources indicated a clustering of 
244 objects centered on a bright $F814W$ = 16.7 elliptical galaxy 
at pixel location $x$ = 789.9, $y$ = 750.9 (see Figure~\ref{fig:image}).  
We reduced the initial catalogue by 6 
objects that look extended, another 5 objects that are very bright 
and red, and 38 objects that were only detected in one filter and therefore 
for which we have no color information.  This final sample of 195 objects 
must be GCs themselves, in orbit around the background 
galaxy.  The clusters are among the faintest objects in our catalogue, 
ranging in brightness from 26 $\lesssim F814W \lesssim$ 30.  Given 
the ACS pixel scale and distance to the galaxy (see below), these 
clusters are all located within a radial distance of 1.7 -- 20~kpc 
(0.85$''$ -- 10$''$) from the center of this galaxy.  The 
incompleteness in our sample increases rapidly as we approach the center of 
the galaxy and therefore we preferentially detect the bluer clusters.  We can roughly 
estimate the masses of the clusters assuming they are old (12~Gyr) and 
metal-poor ([Fe/H] = $-$1.5).  For a Kroupa initial-mass function, the 
\cite{maraston05} models indicate that the GCs with $F814W = 27.5$ 
are $\sim$2 $\times$ 10$^6$~$M_\odot$ whereas the brightest globulars in 
this galaxy are $\sim$10$^7$~$M_\odot$.

To measure the distance to the galaxy accurately, we first obtained a ground 
based image of the field in excellent seeing conditions and low air-mass with 
the Gemini Multi-Object 
Spectrograph (GMOS) on Gemini South (Program ID: GS-2006A-DD-16, executed on 
18~Aug.\ 2006).  Despite the confusion from NGC~6397 stars, the elliptical 
galaxy was easily delineated at $\alpha_{\rm J2000}$ = 17:41:12.2, 
$\delta_{\rm J2000}$ = $-$53:43:16.5 in a 120~s exposure with the $g'$ filter.  
Next, we obtained a low resolution spectrum of the galaxy with a 1$''$ longslit 
and the B600 grating ($R$ = 1600) centered at 5200~${\rm \AA}$.  The position 
angle was set to 325~degrees to avoid contamination in the slit from nearby 
stars.  Four 900~s target exposures, as well as observations of a flux 
standard, a Cu/Ar arc, and a flat field were obtained.  The data were reduced 
using the Gemini IRAF Package, Version 1.8.  

The final reduced spectrum of the elliptical galaxy in the observed wavelength 
frame of reference is shown in Figure~\ref{fig:cmd} (top).  The arrows mark 
several well-defined spectral features along with their rest-frame 
wavelengths.. A cross-correlation with a template spectrum of 
an old, metal-rich single stellar population yields a best-fit 
redshift of $z = 0.0894$. Assuming standard concordance 
cosmology with $\Omega_{m} = 0.3$, $\Omega_{\Lambda} = 0.7$, 
and $h = 0.7$, this redshift corresponds to a proper radial 
distance of 375~Mpc and a lookback time of 1.2~Gyr. The luminosity 
distance of 409~Mpc gives a distance modulus of $m-M = 38.06$.  
While this is not the most distant galaxy in which GCs 
have been detected (that currently belongs to Abell 1689 at $z=0.183$; 
Mieske et~al.\ 2004), the photometry in this galaxy is considerably 
deeper than in any previous study of a galaxy at a comparable distance.

\section{Results -- Photometric Properties of \\ Distant Globular Clusters} \label{results}

To probe the properties of these distant GCs, we compare 
their color distribution to that of globulars in local elliptical 
galaxies.  These galaxies nearly always have a bimodal distribution of 
GC colors, with peaks near $V-I \sim 0.95$ and 1.18.  There 
is a mild dependence of the peak GC colors with host 
galaxy luminosity (Strader, Brodie, \& Forbes 2004; Peng et~al.\ 2006), 
and so we correct these values to $V-I \sim 0.93$ and 1.16 for a galaxy 
with $M_{\rm V}$ = $-$20.5 (assuming $V-I$ = 1.2).  We note that our 
measurement of the integrated brightness of the galaxy is only roughly 
correct given the large number of foreground contaminating stars in the 
outer parts.  This does not effect our results as a 0.5 magnitude 
difference in the galaxy luminosity would translate to a $<$0.01 mean 
magnitude offset in the expected peak color of the GCs.  

\begin{figure}
\epsscale{1.2} \plotone{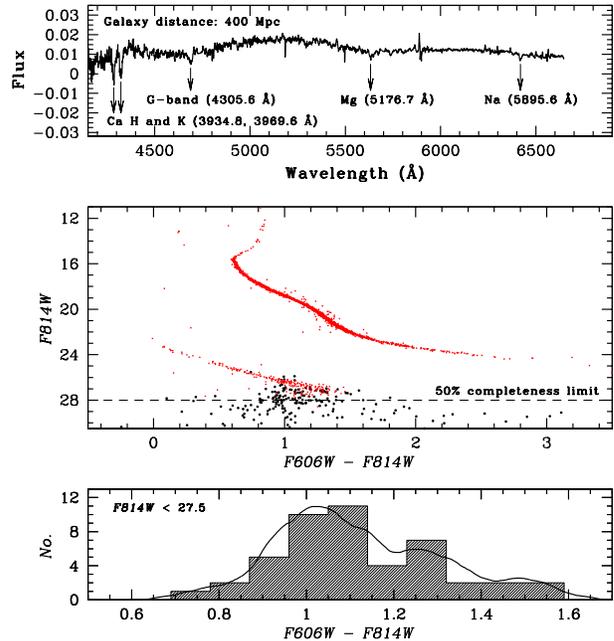} \figcaption{{\it Top} -- A Gemini GMOS 
spectrum of the elliptical galaxy shown in Figure~\ref{fig:image} 
reveals its redshift to be $z = 0.0894$, indicating a luminosity 
distance of 409~Mpc.  Several absorption features and their rest-frame 
wavelengths are illustrated. {\it Middle} -- The proper motion selected 
color-magnitude diagram of the foreground cluster NGC~6397 shows a 
beautiful main-sequence extending from the turnoff down to the lowest 
mass stars that burn hydrogen (red points).  The rich white dwarf cooling 
sequence of the cluster is also visible.  The larger dots in the 
faint-blue part of the CMD represent the 195 extragalactic GCs found 
in the vicinity of the large elliptical galaxy in 
Figure~\ref{fig:image}.  {\it Bottom} -- Histogram and 
density estimate for the colors of 46 GCs with $F814W < 27.5$.  
The majority of GCs belong to the blue subpopulation, with a tail 
towards red objects. \label{fig:cmd}}
\end{figure}

In Figure~\ref{fig:cmd} (middle) we present the color magnitude diagram of 196 
GCs.  This includes the foreground globular NGC~6397 where 
we resolve all of the individual stars ({\it small points}) as well as the 
195 extragalactic objects ({\it larger points below the white dwarf cooling 
sequence of NGC~6397}).  A large fraction of these distant GCs 
($\sim$60\%) have very faint magnitudes and lie below the 50\% completeness limit 
({\it dashed line}, $F814W$ = 28).  A total of 46 objects, or $\sim$64 
after accounting for incompleteness, have $F814W < 27.5$.  The foreground 
reddening in the direction of NGC~6397 is E($F606W - F814W$) = 0.18 
\citep{hansen07}, giving $A_{F814W}$ = 0.33 for a standard reddening curve. 
Using the distance modulus in \S\,2 and this reddening, the expected peak 
of the approximately lognormal GC luminosity function (GCLF) 
is at $F814W$ $\sim$ 30.0. It follows that we have only observed the 
tip of the GCLF in this galaxy. Assuming a standard $\sigma = 1.3$ for the 
GCLF (e.g., Kundu \& Whitmore 2001), down to $F814W = 27.5$ only $\sim$2.7\% 
of the total GC population is visible (assuming a Gaussian form
for the entire GCLF). Even though this estimate is 
very rough, it appears that there must be more than a thousand 
GCs in this galaxy. This puts the galaxy squarely in 
the regime of massive ellipticals, and we can safely use the peak 
GC colors of local ellipticals as a reference.

In Figure~\ref{fig:cmd} (bottom) we show a color histogram of GC 
candidates with $F814W < 27.5$ and a density estimate overplotted. There 
is a broad peak of blue objects with $F606W - F814W$ between $\sim$0.8 
and 1.2, and a tail to redder values. Mixture modeling of the color 
distribution with Nmix (see discussion in Strader et~al.\ 2006) finds 
that the preferred fit is two Gaussians centered at $F606W - F814W$ = 
1.03 $\pm$ 0.02 and 1.29 $\pm$ 0.05. The errors are derived through 
bootstrapping.  While increasing photometric errors 
tend to broaden the color distribution as one goes to fainter magnitudes, 
performing the same fitting to a limit of $F814W = 28$ yields essentially 
the same result: a preference for two populations of GCs, 
with peaks at $F606W - F814W$ = 1.03 and 1.31. The small number of red 
GCs makes that peak value uncertain, but the derived blue 
peak appears consistent with a visual estimate from Figure~\ref{fig:cmd} 
(bottom).  


We now compare these observed peak values to those expected from a 
scaling of local GC systems in a galaxy with same luminosity 
as our host elliptical.  To convert from $V-I$ to $F606W-F814W$, we use the 
empirical relation of \cite{harris07}, derived from photometry of NGC~2419. This 
relation is $F606W - F814W = 0.707 (V-I) + 0.033 (V-I)^{2}$. To these values we 
add Galactic reddening of E($F606W - F814W$) = $0.18 \pm 0.03$. Finally, we add a 
$k$-correction derived from Bruzual \& Charlot (2003) stellar 
population models. This depends slightly on the (unknown) metallicity 
of the stellar population, but assuming values of [Fe/H] $\sim -1.2$ 
and $-0.2$ for the two subpopulations gives $k$-corrections of 0.088 
and 0.063, respectively. If our assumed metallicities are incorrect, 
the effect on the $F606W - F814W$ colors is minor: a 0.1 dex change in 
[Fe/H] corresponds to about 0.01 mag, and 0.2 dex is a reasonable 
assumption for the uncertainty in the metallicity. Combining these 
corrections together gives predicted peak values of $F606W - F814W = 
0.95 \pm 0.03$ and $1.11 \pm 0.04$ \emph{disregarding} evolution in 
the stellar population.  We also computed the predicted peak colors 
assuming the \cite{sirianni05} transformations between Vega magnitudes 
and ACS filters and found the same results, within the small error 
bars.

The observed change in the peak colors of blue and red GCs from 
$z = 0.089$ to today is 0.08 $\pm$ 0.04 and 0.18 
$\pm$ 0.06 magnitudes. What is the expected evolution?  Both the 
\cite{maraston05} and \cite{bruzual03} models predict that the red metal-rich 
GCs should be about 0.01 mag redder in $F606W - F814W$ at 
the present day, and that there should be essentially \emph{no} 
($< 0.01$ mag) evolution in the blue metal-poor GCs.  
Therefore, to first order, our results suggest that more evolution has 
occurred than is predicted by the models, especially for the red 
clusters\footnote{Based on the tightness of NGC 6397's main-sequence, the differential 
reddening along this line of sight is negligible.}.  We note 
that some GC systems in the local universe 
show the peculiar property of a correlation between blue GC 
magnitude and color, such that the more luminous GCs are 
redder \citep{harris06,strader06,mieske06}. Since we sample 
only the brightest GCs in this distant galaxy, the mean color 
of the blue GCs derived above may be redder than the typical 
object, therefore bringing the predicted and observed values closer to 
the expected difference.  We therefore conclude that the behavior of the 
blue GCs is essentially consistent with simple stellar population 
models, while the red GCs are substantially redder at a lookback 
time of $\sim 1.2$ Gyr than expected.

\section{Discussion}

The most obvious candidate for any change in blue GC colors 
is the horizontal branch (HB). At old ages, its morphology is highly sensitive 
to a number of variables, including age.  At younger ages, GCs 
should typically have red HBs, which transition to blue HBs as the stellar 
population ages.  For old systems, the quantitative evolution in a relatively 
red color such as $F606W - F814W$ is expected to be small.  Our results confirm 
this, suggesting that blue GCs have become bluer by just 
0.08 $\pm$ 0.04 magnitudes over the past 1.2~Gyr.  The \cite{maraston05} 
models for a 12--13 Gyr metal-poor ([$Z$/H] = $-1.35$) single stellar population 
predict a difference of only 0.02 mag between an object with a blue and a red HB, 
consistent with our result at 1.5$\sigma$.  Observations of Galactic GCs show 
no correlation between residuals 
of $V - I$ colors from a fiducial $V - I$ to [Fe/H] relation and two different 
parameterizations of HB morphology, suggesting that---at least for the range of HBs 
observed in the Galaxy---the HB has little effect on the $V - I$ color 
(Smith \& Strader 2007).

For the red GCs, our mixture modeling fit of the color distribution 
indicates a peak at $F606W - F814W$ = 1.29 $\pm$ 0.05, 0.18 $\pm$ 0.06 
magnitudes redder than the local sample.  Although our data set contains few red 
GCs, the observed peak is found to be at the same color irrespective 
of the magnitude limit adopted (i.e., pushing the sample to $F814W <$ 28).  
Over the past 1.2~Gyr, the expected evolution from models of a single stellar 
population is much smaller than this, as noted above.  One possible way to 
reconcile {\it some} of these observed differences is to consider whether a fraction 
of these clusters are significantly younger than 12~Gyr, and more metal-rich.  
In this case, the predicted color difference resulting from standard stellar 
evolution between a redshift of $z = 0.089$ and now would be larger than that 
calculated above.  

An additional effect may explain the observed color difference if the metallicities 
of the clusters are in fact supersolar.  Recently, \cite{kalirai07b} have shown 
that the remnant population of white dwarfs in the old (8~Gyr), supersolar metallicity 
Galactic star cluster NGC~6791 are undermassive when compared to more metal-poor 
systems.  These stars likely formed through a unique channel involving enhanced 
mass loss of the progenitor star on the red giant branch.  Such evolution naturally 
explains both the absence of stars near the tip of the red giant branch in this 
cluster and the population of the extreme horizontal branch of the system (see 
also Castellani \& Castellani 1993).  Although the morphology of the horizontal branch does not 
heavily affect the color evolution of a population in our filters, the depletion 
of stars near the tip of the red giant branch can be more important and lead to 
an increasingly bluer population as the system ages.  Of course, this mechanism 
would be even more efficient if the clusters are older since the envelope mass 
of the evolving stars is lower.  However, this mechanism does require the 
clusters to have [Fe/H] $\gtrsim$ $+$0.2 \citep{kilic07}, and we see few systems 
nearby that are this metal-rich.

Finally, we note that these objects are far enough away that we cannot rule out 
a contaminating population of ultra-compact dwarfs (UCDs).  The properties of 
these potentially transition objects (between GCs and dwarf galaxies) 
vary widely among galaxies.  For example, in NGC~3311, they are are seen as very 
red objects more luminous than the old and red GC population 
(Wehner \& Harris 2007).  However, in NGC~1407, the UCDs have larger sizes than 
globulars but have colors that fall between the blue and red subpopulations 
(Harris et~al.\ 2006).  We note that UCDs are typically found in massive galaxy 
clusters or ``groups'', and it is difficult to assess the environment near our 
elliptical galaxy given the presence of NGC~6397 stars out to several arcminutes 
(1$'$ = 120~kpc).

\section{Conclusions} \label{conclusions}

We have presented a study of the photometric properties of one of the most 
distant sample of GCs yet discovered, 195 objects at $d$ = 400~Mpc.  
These clusters are all found around a bright elliptical galaxy 
that is itself located behind one of the nearest Galactic GCs 
to the Sun, NGC~6397.  Given their large distance, the light from these 
clusters has taken 1.2~Gyr to reach us, and therefore these objects 
offer us a rare chance to probe the evolution of GCs in 
the recent past.  By comparing the color function of these clusters to nearby 
GCs, we find differences that suggest the mean colors of GCs have become 
bluer in the past Gyr.  This evolutionary change can be 
explained assuming standard stellar evolution for the blue clusters.  However, 
we find differences that suggest the mean colors of the red GCs 
were significantly redder 1.2~Gyr ago as compared to samples in local giant 
ellipticals, much redder than evolutionary models predict. 

\acknowledgements
We would like to thank the entire {\it HST}/ACS team from GO-10424 for 
their help in obtaining the imaging observations.  We are also grateful 
to the Gemini South observatory for approving our program under the Director's 
Discretionary Time allotment (Program ID: GS-2006A-DD-16).  We would also 
like to thank an anonymous referee for several useful suggestions that have 
improved the quality of this Letter, and for his/her patience through the 
refereeing process.  JSK and JS are both supported by NASA through Hubble Fellowship 
grants, awarded by the Space Telescope Science Institute, which is operated by the 
Association of Universities for Research in Astronomy, Incorporated, under NASA 
contract NAS5-26555. Support for this work was also provided by grant 
HST-GO-10424 from NASA/STScI.  The research of HBR is supported 
by grants from the Natural Sciences and Engineering Research Council of Canada.  
He also thanks the Canada-US Fulbright Program for the award of a Fulbright 
Fellowship.

\end{document}